\documentclass[pre,aps,showpacs,preprint,floatfix,amsmath,amssymb]{revtex4}

\usepackage{amsmath,amssymb}
\usepackage{graphicx}
\usepackage{psfrag}
\usepackage{graphicx}
\usepackage{epsf}

\begin{document}
\title {Universal properties of
three-dimensional magnetohydrodynamic
turbulence: Do Alfv\'en waves matter?}

\author
{Abhik Basu$^1$\cite{byemail} and Jayanta K Bhattacharjee$^2$}

\affiliation{
$^1$Max-Planck-Institut f\"{u}r
Physik Komplexer Systeme, N\"{o}thnitzer Strasse 38, D-01187, Dresden, Germany,
and Poornaprajna Institute of Scientific Research, Bangalore, India\\
$^2$Department of Theoretical Physics, Indian Association for the
Cultivation of Sciences, Jadavpur, Calcutta 700032, INDIA.}
\date{\today}

\begin{abstract}
We analyse the effects of the propagating Alfv\'en waves, arising
due to non-zero mean magnetic fields, on the nonequilibrium steady
states of three-dimensional ($3d$) homogeneous Magnetohydrodynamic
(MHD) turbulence. In particular, the effects of Alfv\'en waves on
the universal properties of
$3d$MHD turbulence are
studied in a one-loop self-consistent mode-coupling approach.
We calculate the kinetic- and magnetic energy-spectra.
We find that {\em even} in the presence of a mean magnetic
field the energy spectra are Kolmogorov-like, i.e., scale as $k^{-5/3}$ in the
inertial range where $\bf k$ is a Fourier wavevector belonging to the inertial
range. We also elucidate the
multiscaling of the structure functions in a log-normal model by evaluating the
relevant intermittency exponents, and our results suggest
that the multiscaling deviations from the simple Kolmogorov scaling of the
structure functions decrease with increasing strength of the mean
magnetic field. Our results compare favourably with many existing
numerical and observational results.

\end{abstract}


 \maketitle

\section{Introduction}
The effects of propagating waves on the statistical properties of systems out
of equilibrium remain an important topic of discussion. In the context of a
coupled spin model in one-dimension $(d=1)$\cite{dib} it has been shown that
the presence of such waves leads to weak dynamic scaling in that model. In
contrast, in a coupled Burgers-like model in $d=1$ propagating waves do not
affect the scaling properties of the correlation functions at all \cite{abjkb}.
So far such issues have been considered only within very simplified
one-dimensional nonequilibrium models \cite{abjkb,verma}. Magnetohydrodynamic
($3d$MHD) turbulence, which is a hydrodynamic description of the coupled
evolution of the velocity fields $\bf u$ and the magnetic fields $\bf b$ in a
quasi-neutral plasma, stands as a very good candidate for a natural system with
propagating waves in three dimensions as most of its natural realizations have
proagating Alfv\'en waves arising due to the presence of mean magnetic fields.
Examples of such physical situations include solar wind, neutral plasma in
fusion confinement devices etc. The presence of the propagating Alfv\'en modes,
in addition to the usual dissipative modes due to the fluid and magnetic
viscosities makes MHD turbulence a good natural example to study the interplay
between the propagating and the dissipative modes in a system and their
combined effects on the scaling properties of the correlation and structure
functions, which are important issues from the point of view of nonequilibrium
statistical mechanics.

The scaling of magnetic- [$E_b(k)$] and kinetic- [$E_u(k)$] energy
spectra in the inertial range (i.e., wavevector $k$ lies in the
region $L^{-1}\ll k\ll \eta_D^{-1}$, $L$ and $\eta_D$ being the
integral scale given by the system size and the dissipation
scale, respectively) in $3d$MHD in the presence of Alfv\'en waves
originating due to a mean magnetic field $\bf B_o$ remains controversial
till the date. Numerical simulations, due to the lack of sufficient
resolutions failed to conclusively distinguish between the
Kolmogorov's and Kraichnan's predictions (see below). In this
communication, within one-loop self-consistent mode-coupling (SCMC)
approximations, we obtain the following results for homogeneous but anisotropic (due to the mean magnetic field) $3d$MHD:
\begin{itemize}
\item The bare
Alfv\'en wave speed, proportional to
$B_o$, renormalizes to acquire a singular $k$-dependence to
become $\sim B_ok^{-1/3}$ where $k$ is a Fourier wavevector
belonging to the inertial (scaling) range. From this result we
are able to conclude that {\em even} in the presence of a mean
magnetic field the kinetic- and the magnetic- energy spectra scale
as $k^{-5/3}$ in the inertial range, identical with the situation without any
mean magnetic field.
\item The dimensionless Kolmogorov's constants for the kinetic- and
magnetic-energy spectra depends on a dimensionless parameter $\beta$
which we identify as the ratio of the renormalised Alfv\'en wave
speed and the renormalised viscosities (see below).
\item The intermittency exponents of the Els\"{a}sser fields $\bf
z^{\pm}=u\pm b$
which approximately and qualitatively
characterise the multiscaling properties of the structure functions
in a log-normal model depend on the parameter $\beta$ and decrease with
increasing $\beta$.
\end{itemize}

Thus our results
show that Alfv\'en waves in $3d$MHD do not affect the scaling
properties of the two-point correlation functions. However, the multiscaling
properties
of the structure functions are shown to be affected by the mean magnetic fields.

The famous Kolmogorov's arguments \cite{k41} for fluid turbulence
can be easily extended to $3d$MHD turbulence. In MHD, in the unit
where mass density $\rho=1$, the kinetic energy dissipation rate
per unit mass ($\epsilon_K$) and the magnetic energy dissipation
rate per unit mass ($\epsilon_M$) have same physical dimensions.
Thus, as in fluid turbulence, by claiming that the structure
functions of the velocity and magnetic field differences in the
inertial range must be constructed out of the mean energy dissipation
rate $\epsilon$ per unit mass
and the local length scale $r$ (belonging to the
inertial range) one obtains for the $n$-th order structure
function \cite{k41mhd,abthesis}
\begin{equation}
S_n^a(r) \equiv \langle [a_i({\bf x+r})-a_i({\bf x})]\hat
r_i]^n\rangle \sim (\epsilon r)^{n/3},\;\;\eta_D\ll r\ll L
\end{equation}
where $\epsilon=\epsilon_K \, {\rm or}\, \epsilon_M$ and $a=u,b$ for
the velocity and the magnetic fields, $\eta_D$ is the (small) dissipation
scale and $L$ is the (large) system size. This yields, for the energy
spectra in the inertial range (as a function of wavevector $k$)
\begin{equation}
E_a(k)\sim k^{-5/3},\,a=u,b.
\end{equation}
This is known as the K41 theory in the relevant literature. However, in the
presence of a mean magnetic field $\bf B_o$, the Alfv\'en waves are generated
(see below) with the propagation speed $\sim B_o$. Thus in such a case in
addition to the usual Kolmogorov time scale $\sim k^{2/3}$ \cite{yakhot,abepl},
there exists another time scale constructed out of the mean magnetic field
$B_o$, known as the Kraichnan time scale $\sim (B_ok)^{-1}$ \cite{krai}.
Kraichnan argued that this time scale would determine the energy cascade
process and hence would enter in the expression of the structure function
yielding $E_a(k)\sim k^{-3/2},\,a=u,b$. There has not been any satisfactory
resolution of this issue till this date; due to the particularly difficult
vectorial nature of the $3d$MHD equations (see below) it is rather difficult to
achieve high Reynolds number in Direct Numerical Solutions (DNS) of the $3d$MHD
equations. Numerical solutions of an MHD shell-model in the presence of a small
mean magnetic field-like term did not find any dependence of the multiscaling
of the structure functions on the mean magnetic field \cite{abthesis}.
Analytically, it has been shown within the context of a $1d$ coupled
Burgers-like model \cite{abjkb} that the energy spectra are independent of a
mean magnetic field and in case of $1d$ MHD turbulence it exhibits the
Kolmogorov scaling for the energy spectra. Similar conclusions followed from an
analogous one-dimensional model \cite{verma}. Various phenomenological
approaches, including weak turbulence theories and three-wave interaction
models, yield, in general, mean magnetic field dependences of the energy
spectra \cite{galtier,ng,montgomery,shebalin} when the mean magnetic field is
strong. However, these theories, despite their predictions are either not
directly derivable from the underlying $3d$MHD Eqs. of motion or involve
additional assumptions on the flow fields. Analyses starting from the $3d$MHD
Eqs. in this regard are still lacking. Most observational results on
astrophysical systems seem to favour the K41 results \cite{obser}. Simulations
of incompressible MHD \cite{simu1} find results close to the K41 result.
Simulations of compressible MHD \cite{simu2}, even though in some cases find
energy spectra closer to K41, in general yields a less clear picture. Recent
numerical results of M\"uller {\em et al} \cite{biskampnew} suggest that in
presence of finite magnetic helicity structure functions parallel and
perpendicular to the mean magnetic fields are affected differently by the mean
magnetic field. In this paper, we address some of these issues by starting from
the $3d$MHD Eqs. without making any further assumptions on the velocity and the
magnetic fields, except for the validity of the perturbative approaches. In
particular, we show by applying one-loop mode coupling methods on the $3d$MHD
equations with a mean magnetic field and in the absence of any magnetic
helicity that the one-dimensional energy spectra in $3d$MHD turbulence are
independent of the mean magnetic field $\bf B_o$ and scale as $k^{-5/3}$ in the
inertial range where $\bf k$ is a Fourier wavevector belonging to the inertial
range. We then proceed to calculate the Kolmogorov's constants for the kinetic-
and the magnetic-energy spectra and show that they depend on $B_o$. Lastly, we
calculate the intermittency exponents for the velocity and the magnetic fields
and find that they decrease with increasing $B_o$. We do not distinguish
between the longitudinal and the transverse structure functions. Even though
for analytical convenience we assume a weak mean magnetic field, we are able to
obtain new and interesting results concerning scaling and multiscaling in
$3d$MHD in the presence of a mean magnetic field of small magnitude. In this
respect our results can be considered as complementary to some of the existing
results \cite{galtier,montgomery,ng}. The rest of the paper is organised as
follows: In Sec.\ref{model} we discuss the $3d$MHD equations for incompressible
fluids. In Sec.\ref{incom} we show that for incompressible fluids the bare
Alfv\'en wave speed $\sim B_o$ renormalizes to pick a correction $k^{-1/3}$ in
the inertial range. This, as we argue, implies that the energy spectra scale as
$k^{-5/3}$ in the inertial range {\em even} in the presence of a mean magnetic
field. In Sec.\ref{kolcons} we calculate the Kolmogorov's constants for the
kinetic and the magnetic energy spectra. We introduce a parameter $\beta$ which
is the dimensionless ratio of the renormalised mean magnetic field and
renormalised viscosities and show that the Kolmogorov's constants depend on
$\beta$. In Sec.\ref{inter} we elucidate the multiscaling properties of the
structure functions by calculating the intermittency exponents and show that
they decrease with increasing $\beta$, i.e., with increasing $B_o$. This
suggests that a mean magnetic field tends to reduce multiscaling corrections to
the K41 results for the structure functions. In Sec.\ref{conclu} we summarize
our results.

\section{Model equations}
\label{model}

We begin by writing down the $3d$MHD equations for the velocity fields 
$\bf u$ and the magnetic fields $\bf b$: The velocity field $\bf u$
is governed by the Navier-Stokes equation modified by the
inclusion of the Lorentz force \cite{jack}
\begin{equation}
\frac{\partial {\bf u}}{\partial t}+({\bf u}\cdot\nabla ){\bf u}=
- \frac{\nabla p}{\rho}+\frac{(\nabla \times {\bf b})\times \bf b}{\rho}+
\nu\nabla^2 {\bf u}+{\bf f},
\label{mhdvcom}
\end{equation}
and the dynamics of the magnetic field $\bf b$ is governed by the
Induction equation \cite{jack} constructed out of the Ohm's law
for a moving frame and the Ampere's law:
\begin{equation}
\frac{\partial \bf b}{\partial t}=\nabla\times ({\bf u\times b})
+\eta \nabla^2 {\bf b}+{\bf g}. \label{mhdb}
\end{equation}
Here, $\rho$
is the mass density, $p$ is the pressure, $\nu\;$ is the fluid
viscosity and $\eta$ is the magnetic viscosity
(inversely proportional to the electrical conductivity of the
fluid medium concerned). Functions $\bf f$ and $\bf g$ are external
forces
needed to maintain a statistical steady state. In the present
approach these are taken to be stochastic forces. We assume them
to be zero-mean and Gaussian distributed with specified variances
(see below). In addition to Eqs. (\ref{mhdvcom}) and (\ref{mhdb})
we also have $\bf \nabla \cdot b=0$ (Maxwell's equation) and, for
incompressible fluids $\bf \nabla\cdot u=0$.

If the magnetic fields ${\bf b}({\bf x}, t)$ are such that
$\langle  {\bf b}\rangle = {\bf B_o}$ (a constant vector)
then replacing $\bf b$ by
$\bf b+B_o$ where now $\langle \bf b\rangle=0$, in Eqs. (\ref{mhdvcom}) and
(\ref{mhdb})
one obtains additional linear terms proportional to wavevector
$\bf k$ leading to wave-like excitations, known as Alfv\'en waves.
The resulting equations are
\begin{equation}
\frac{\partial {\bf u}}{\partial t}+\lambda_1({\bf u}\cdot\nabla ){\bf u})=
- \frac{\nabla p}{\rho}+\lambda_2\frac{(\nabla \times {\bf b})\times {\bf
b}}{\rho}+\frac{(\nabla \times {\bf b})\times {\bf B_o}}{\rho}+
\nu\nabla^2 {\bf u}+{\bf f},
\label{mhdfullv}
\end{equation}
and
\begin{equation}
\frac{\partial \bf b}{\partial t}=\lambda_3\nabla\times ({\bf u\times
b})+\nabla\times ({\bf u\times B_o}) +\eta \nabla^2 {\bf b}+{\bf
g}. \label{mhdfullb}
\end{equation}
with $\langle \bf b ({\bf x},t)\rangle =0$. The parameters
$\lambda_1,\,\lambda_2,\,\lambda_3$ are kept for book keeping purposes and can
be set to unity (see Sec.\ref{symm}). Note that on dropping
the nonlinear and the dissipative terms from Eqs. (\ref{mhdfullv}) and
(\ref{mhdfullb}) the resulting linear
coupled partial differential equations admit wave-like solutions
with dispersion relation linear in wavevector $k$. These are known
as the Alfv\'en waves \cite{mont} in the literature which propagate with speed
proportional to $B_o$. In the
following sections we would calculate the kinetic and the magnetic
energy spectra, the Kolmogorov's constants and the intermittency
exponents
in the presence of the Alfv\'en waves, i.e., for $B_o\neq 0$.

\section{Correlation and structure functions in MHD}
\label{corr}

In the statistical steady state the time dependent correlation
functions of $\bf u$ and $\bf b$ exhibit scaling which are
characterized by the roughness exponents $\chi_u$ and $\chi_b$ respectively,
and the dynamic exponent
$z$. In terms of the scaling exponents
$\chi_u,\,\chi_b$ and $z$ the velocity and the
magnetic field correlators have the form (as a function of wavevector $\bf k$
and frequency $\omega$)
\begin{eqnarray}
C^u_{ij}(k,\omega)=\langle u_i({\bf k},\omega) u_j({\bf
-k},-\omega)\rangle=D^uP_{ij}k^{-d-2\chi_u-z}f_u(k^z/\omega),
\nonumber
\\
C^b_{ij}(k,\omega)=\langle b_i({\bf k}, \omega)b_j({\bf
-k},-\omega)\rangle=D^bP_{ij}k^{-d-2\chi_b-z}f_b(k^z/\omega),
\label{corrform}
\end{eqnarray}
where $f_u$ and $f_b$ are scaling functions, $P_{ij}$
is the transverse projection operator:
$P_{ij}=\delta_{ij}-k_ik_j/k^2$ to account for the incompressibility of the
fields. The kinetic-
and magnetic-energy are simply related to the correlation
functions: They are just the equal time velocity and magnetic
field auto correlation functions multiplied by appropriate
phasefactors. We now set out to find whether the Alfv\'en waves
are relevant perturbations on the system. It should be noted that the
correlators (\ref{corrform}) are chosen as they would be in the fully isotropic
case. In the presence of a mean magnetic field $\bf B_o$ there would however be
additional anisotropic terms in the expression for the correlators above. In
the small $\bf B_o$ limit the lowest order corrections to the expressions
(\ref{corrform}) are $O(B_o)^2$. Neglecting these correction terms does not
affect our scaling analyses below, and in any case, we are interested to see
whether anisotropic Alfv\'en waves are relevant perturbations in the large
scale, long time limit on the isotropic $3d$MHD as characterised by expressions
(\ref{corrform}). Therefore, it suffices for us to work with the expressions
(\ref{corrform}) for our mode-coupling analyses below.

A complete characterisation of the nonequilibrium steady state (NESS)
of MHD, however, requires informations about, not just the
scaling of the energy-spectra, but also of the $n$-th order equal time structure functions of the
velocity and the magnetic fields in the inertial range. These are defined by
\begin{equation}
S_n^a(r)\equiv\langle|[a_i({\bf x+r})-a_i({\bf x})]\hat r_i|^n\rangle,
\;\;a=u,b,\,\eta_D\ll r\ll L,
\label{defstru}
\end{equation}
which scale as $r^{\zeta_n^a}$ where $r$ belongs to
the inertial range ($\eta_D\ll r\ll L$).
According to the Kolmogorov theory (K41)
the multiscaling exponents $\zeta_n^a=n/3$; i.e., they are linear in
$n$. Subsequent numerical and observational results
for homogeneous and isotropic $3d$MHD, i.e., without any mean magnetic fields, \cite{abprl,biskamp3d}
suggested deviations from the K41 results which are similar to those of pure fluid turbulence \cite{rahulrev}.
Much less results for the multiscaling of the structure functions are available when there are mean magnetic fields.
Recent numerical simulations of M\"uller {\em et al} \cite{biskampnew} suggested that structure functions parallel and
perpendicular to the mean magnetic fields are differently affected by it when
there is a finite magnetic helicity. Below we investigate the issue of
multiscaling in presence of a mean magnetic field $\bf B_o$ within the context
of a log-normal model by evaluating the relevant intermittency exponents which
are found to depend explicitly on $B_o$.
In our analyses we ignore the distinction between the structure functions
parallel and perpendicular to the direction of the mean magnetic field;
nevertheless, as we discuss below, our
results are significantly new.

\section{Symmetries of the equations of motion}
\label{symm}

We begin by re-expressing $3d$MHD Eqs. (\ref{mhdvcom}) and (\ref{mhdb}) 
by writing
the magnetic fields $\bf b$ as a sum of a space-time dependent part and a
constant vector: ${\bf b}({\bf x},t)\rightarrow {\bf b}({\bf x},t)+{\bf \tilde
B_o}$. In terms of these fields and parameters, the $3d$MHD Eqs. become
\begin{equation}
\frac{\partial {\bf u}}{\partial t}+\lambda_1({\bf u}\cdot\nabla ){\bf u})= -
\frac{\nabla p}{\rho}+\lambda_2\frac{(\nabla \times {\bf b})\times {\bf
b}}{\rho}+\frac{(\nabla \times {\bf b})\times {\bf \tilde B_o}}{\rho}+
\nu\nabla^2 {\bf u}+{\bf f}, \label{mhdfullv1}
\end{equation}
and
\begin{equation}
\frac{\partial \bf b}{\partial t}=\lambda_3\nabla\times ({\bf u\times
b})+\nabla\times ({\bf u\times \tilde B_o}) +\eta \nabla^2 {\bf b}+{\bf g}.
\label{mhdfullb1}
\end{equation}
It should be noted that in Eqs. (\ref{mhdfullv1}) and (\ref{mhdfullb1}) $\bf
\tilde B_o$ is {\em not} the mean magnetic field in general; it would be so
only if $\langle {\bf b} ({\bf x},t)\rangle $ is zero in Eqs. (\ref{mhdfullv1})
and (\ref{mhdfullb1}). Note that our above way of splitting the magnetic fields
fields does not change the actual mean magnetic field in the
system: It is still given by $ {\bf \tilde B_o}+\langle{\bf b}\rangle={\bf
B_o}$. The Eqs. of motion (\ref{mhdfullv1}) and (\ref{mhdfullb1}) are invariant
under the following continuous transformations \cite{abepl,abpassive}:
\begin{itemize}
\item The Galilean transformation (TI):
${\bf u}({\bf x},t)\rightarrow {\bf u} ({\bf x
+u_o}t,t)+{\bf u_0},\;\frac{\partial}{\partial t}-{\bf u_0}.\nabla,$
and ${\bf
b}\rightarrow {\bf b}$
\cite{abjkb,abepl,fns} with $\lambda_1=\lambda_3=1$ in Eqs. (\ref{mhdfullv}) and
(\ref{mhdfullb}). This implies
non-renormalization of $\lambda_1$ \cite{abjkb,fns,freykpz}.

\item The transformation (TII) ${\bf \tilde B_0\rightarrow \tilde B_0}+\lambda_2\delta$,
${\bf b}({\bf x},t) \rightarrow {\bf b}({\bf x},t)- {\bf \delta}$, $\bf
u\rightarrow u$. This allows one to work with the {\em effective} magnetic
fields defined by $\sqrt \lambda_2 \bf b$ such that the coefficient of the
Lorentz force vertex constructed out of the effective magnetic fields does not
renormalize. This, therefore, ensures $\lambda_2$ can be set to unity
\cite{abepl,abpassive} by treating all magnetic fields as {\em effective
fields}. Here the shift $\delta$ is a vector. A transformation similar to TII
above exists in a problem of passive scalar turbulence \cite{anton}.
\end{itemize}

The transformation TII essentially signifies the freedom to split the total
magnetic fields as a sum of a constant part and a space time dependent part. It
should be noted that the transformation TII keeps the mean magnetic field
unchanged. In fact, nonrenormalization of $\lambda_2$ can be shown in a simpler
way: Let us assume that under mode eliminations and rescaling
$\lambda_2\rightarrow \alpha\lambda_2$. This scale factor of $\alpha$ can now
be absorbed by redefining the units of the magnetic fields by $\sqrt \alpha \bf
b \rightarrow b$. Therefore, $\lambda_2$ can be set to unity if all magnetic
fields are considered as {\em effective} or rescaled magnetic fields. Since the
Induction Eq. (\ref{mhdb} is linear in the magnetic fields $\bf b({\bf x},t)$,
such rescaling leaves every conclusion unchanged. Of course, the external force
$\bf g$ is also scaled by a factor $\sqrt \alpha$ which does not affect out
analysis here as the assignment of
canonical dimensions to various fields and parameters is done after
absorbing $\lambda_2$ in the definition of $\bf b$.
More specifically under the
rescaling ${\bf x} \rightarrow l{\bf x},\,t\rightarrow l^z t,\,u_i\rightarrow
l^{\chi_u}u_i, \, b_i\rightarrow l^{\chi_b}b_i$ the bare parameters scale as
$\lambda_{1,3}(l)\rightarrow l^{\chi_u+z-1} \lambda_{1,3},\,\lambda_2(l)
\rightarrow l^{2\chi_b-\chi_u+z-1}\lambda_2$ and $\tilde B_o \rightarrow
l^{\chi_b-\chi_u+z-1}\tilde B_o$.  Since there are no fluctuation corrections
to the nonlinearities $\lambda_1,\,\lambda_2,\,\lambda_3$, which are the
consequences of the invariances under the transformations TI and TII, they can
be kept invariant under rescaling of space and time as mentioned above leading
to $\chi_u=\chi_b =\chi$ and $\chi+z=1$. Thus, under naive rescaling the bare
parameter $\tilde B_o\sim \tilde B_o l^{z-1}$. Furthermore, the invariance
under the transformation TII and the resulting Ward identity ensure that
different ways of implementing one-loop RG by having different values for
$\langle {\bf b}\rangle $ and $\bf \tilde B_o$ in Eqs. (\ref{mhdfullv1}) and
(\ref{mhdfullb1}) subject to the same bare mean magnetic field $\bf B_o=\tilde
B_o+\langle b\rangle$. Since any renormalisation scheme must respect a freedom
of choice as represented by the transformation TII, respective terms must scale
the same way under the rescaling of space and time \cite{abpassive}. Hence, due
to fluctuation corrections $\tilde B_o (l)\sim l^{\chi}$. Furthermore, since
both $\bf \tilde B_o$ and $\bf b$ have the same scaling dimensions given by the
exponent $\chi$, the physical mean magnetic field in the system ${\bf B_o}={\bf
\tilde B_o}+\langle{\bf b({\bf x},t)}\rangle$, if it receives fluctuation
corrections under mode eliminations, must scale as $B_o(l)\sim l^{\chi}$, or if
there are no fluctuation corrections to it due to some special symmetries (as
in the one-dimensional Burgers-like model for MHD in Ref.\cite{abjkb}) will be
irrelevant (in an RG) sense as it will flow to zero as $l^{z-1}$ (since $z$ for
fully developed turbulence is less than unity). This clearly suggests the
possibility of a renormalisation of the Alfv\'en wave speed, akin to the
renormalisation of the sound speed in compressible fluid turbulence
\cite{jkbsound}. As we will see below, for $3d$MHD there are infra-red singular
fluctuation corrections to the bare mean magnetic field leading it to
renormalise in a way consistent with our predictions from the Ward identities
above. It should be noted that our analyses above is independent of the
strength of the bare value of the mean magnetic field. Although, in the
discussion above we have rescaled space isotropically (i.e., $x,y$ and $z$
coordinates are rescaled the same way) and analyze the scale dependence of the
resulting effective parameters and the fields, it should be noted that such a
rescaling does not imply that the effective parameters have isotropic
structures.

\section{ Energy spectra for incompressible fluids}
\label{incom}

We begin by writing down the $3d$MHD equations in the
incompressible limit (i.e., $\bf \nabla\cdot u=0$). We write down
the equations of motion in $\bf k$-space in terms of the
Els\"{a}sser variables $\bf z^{\pm}=u\pm b$. The equations are (we
take the mean magnetic field $\bf B_o$ to be along the
$\hat z$ direction)
\begin{eqnarray}
\frac{\partial z^+_l({\bf k},t)}{\partial t}-iB_ok_zz^+_i+
iP_{lp}k_s\sum z_s^-({\bf q},t)z_p^+({\bf k-q},t)+\eta_+^ok^2z_l^+
+\eta_-^ok^2z_l^-&=&\theta_l^+({\bf k},t),\nonumber \\
\frac{\partial z^-_l({\bf k},t)}{\partial t}+iB_ok_zz^-_i+
iP_{lp}k_s\sum z_s^+({\bf q},t)z_p^-({\bf
k-q},t)+\eta_+^ok^2z_l^-+\eta_-^ok^2z_l^+&=&\theta_l^-({\bf k},t).
\label{zpmeq}
\end{eqnarray}
The stochastic forces $\theta_l^{\pm}$ are linear combinations of $f_l,\,g_l$,
the tensor $P_{lp}$ is the transverse projection  operator which appears due
to the divergence-free conditions on $\bf u$ and $\bf b$, and $\eta_{\pm}^o=\nu
\pm \eta$.

In the absence of any crosscorrelations between the velocity and the
magnetic fields,  the simplest choice for the noise variances,
consistent with the divergence-free conditions on the velocity and
the magnetic fields, are
\begin{eqnarray}
\langle \theta_l^{\pm}({\bf k},t)\theta_m^{\pm}({\bf -k},t)\rangle = 2P_{lm}D_1k^{-y}\delta (t),\nonumber \\
\langle \theta_l^{\pm}({\bf k},t)\theta_m^{\mp}({\bf -k},t)\rangle
= 2P_{lm}D_2k^{-y}\delta (t). \label{noisecor}
\end{eqnarray}
In Eqs.(\ref{noisecor}) we choose $y>0$. In particular, the choice of $y=d$ in
$d$-space  dimensions ensures that the energy flux in the inertial range is a
constant (up to a weak logarithmic dependence on wavevector $k$) which forms
the basis of the K41 theory \cite{k41}, and as a result, the parameters $D_1$
and $D_2$ pick up dimensions of energy dissipation rate per unit mass. Hence,
we will use $y=d$ in our calculations below.  In experimental realisations of
MHD, external forces act on the large scales. In analytical approaches, such
forces are replaced by stochastic forces with variances given by Eqs.
(\ref{noisecor}) for calculational convenience. It should however be noted that
in numerical solutions of the Navier-Stokes equation driven by stochastic
forces with variances similar to Eqs. (\ref{noisecor}) structure functions of
the velocity fields are shown to exhibit multiscaling similar to the
experimental results \cite{rfnse}. Our preliminary results from the numerical
solutions of the isotropic $3d$MHD equations (i.e., no mean magnetic fields)
yield multiscaling similar to those obtained from the $3d$MHD Eqs. driven by
large-scale forces \cite{rfmhd}. Although no such numerical results exist for
the case with a mean magnetic field, it will presumably be true for such a
situation also.
 In a stochastically driven Langevin description,
correlation functions are proportional to the noise correlations and hence it
is necessary to force the Induction Eq. (\ref{mhdfullb}) stochastically, as is
common in the literature (see, e.g., \cite{abjkb,abepl} and references therein).
For such noises in the absence of any mean magnetic field (i.e., for
the isotropic case) it has been shown by using a one-loop mode
coupling theory that the scaling exponents have values
$\chi=1/3,\,z=2/3$ \cite{abepl}.

The Eqs. of motion (\ref{zpmeq}) are coupled even at the linear level leading to
a bare propagator matrix for the Eqs. (\ref{zpmeq}) of the form (a '$o$' refers
to bare quantities)

\[{G_o}^{{-1}}=\left(
\begin{matrix}{-i\omega -iB_ok_z+\eta_+^o k^2}&
\eta_-^o k^2 \cr {\eta_-^ok^2} & {{-i\omega +iB_ok_z+\eta_+^o k^2}  }
\end{matrix}
\right)
\]

We use a one-loop mode coupling theory which is conveniently formulated
in terms of the self-energy matrix $ \Sigma$ and the correlation
functions given by Eqs. (\ref{corrform}). The self-energy matrix
$\Sigma$ is defined by
$G^{-1}=G^{-1}_o-{ \Sigma}$ where $G$ is the renormalised
propagator matrix. In terms of the scaling exponents $\chi$ and $z$, and the
{\em renormalized} parameters the self-energy matrix $ \Sigma$ is given by

\[{ \Sigma} =\left (
\begin{matrix}
{\-i\omega -iB(k)k_z +\eta_+k^z}&
\eta_-k^z \cr {\eta_-k^z} &
{-i\omega +iB(k)k_z+\eta_+k^z}
\end{matrix}
\right)
\]
where $B(k)$ is the {\em renormalised} or the {\em effective} Alfv\'en wave
speed. If there are diagrammatic corrections to the imaginary parts of $
\Sigma(k,\omega)$ at frequency $\omega=0$ which are singular in the infra-red
limit, then $B(k)$ is different from $B_o$, the bare Alfv\'en wave speed, else
they are the same.

\begin{figure}[h]
\includegraphics[width=4in]{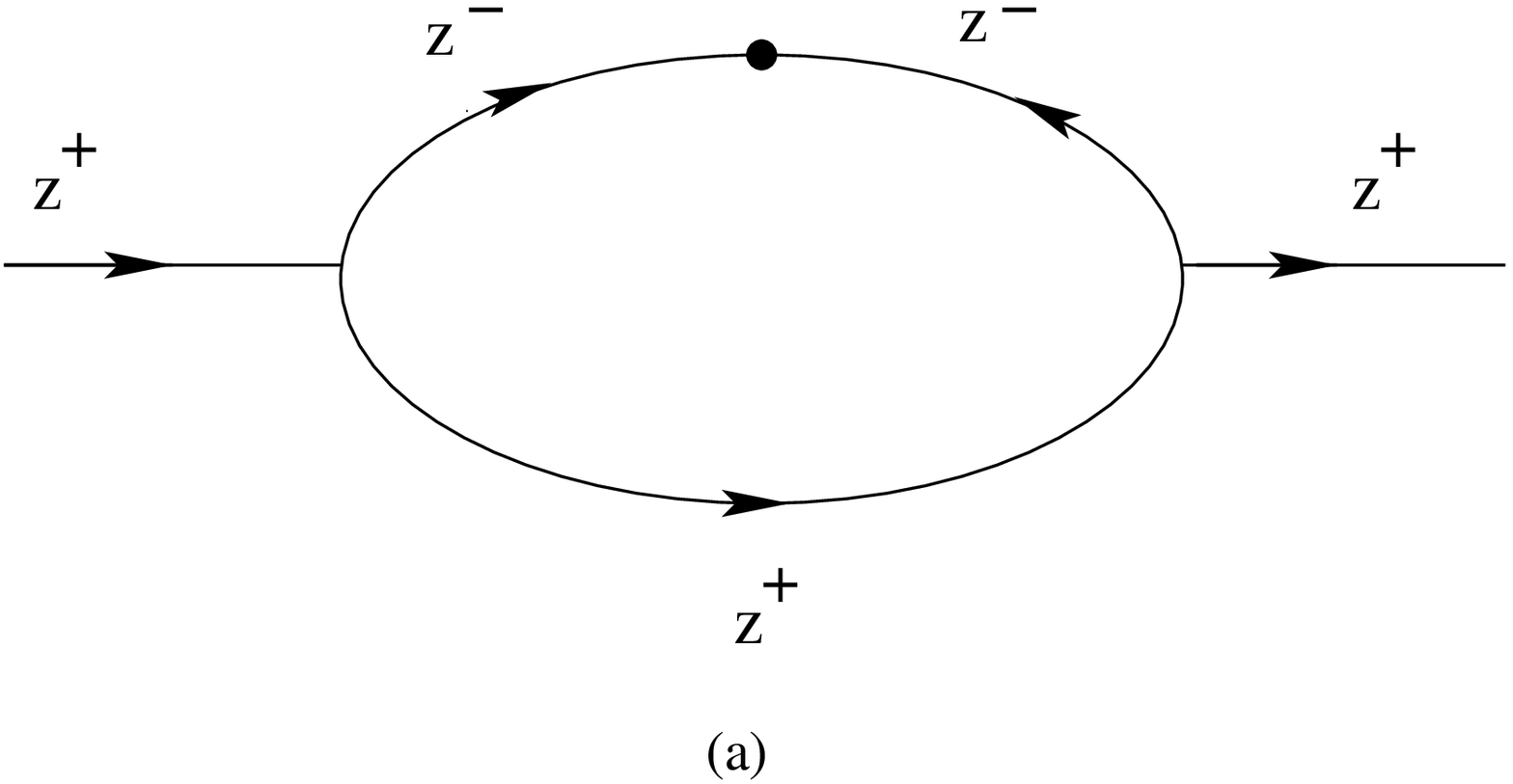}\hfill \includegraphics[width=4in]{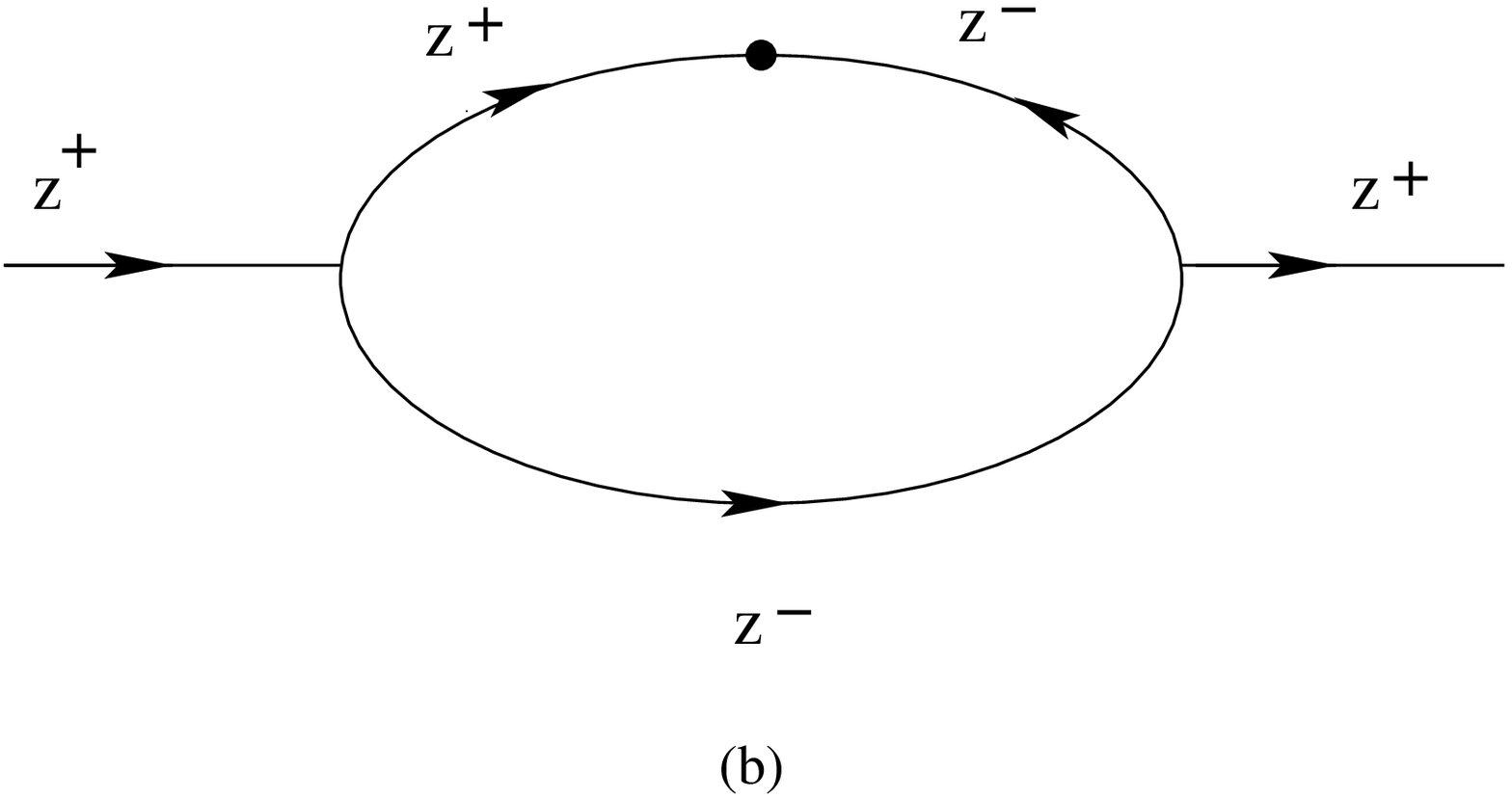}
\caption{One-loop diagrams contributing to renormalization of $G_o^+$.
A single line refers to a propagator and a line with a filled circle refers to a
correlator.}
\label{diag1}
\end{figure}

There are two one-loop diagrams which contribute to the fluctuation corrections
to ${ \Sigma_{11}}(k,\omega=0) =-iB(k)k_z+\eta_+k^z$. These are shown in
Fig.(\ref{diag1}). These have both real and imaginary parts at frequency
$\omega=0$. Thus there are fluctuation corrections to the bare Alfv\'en wave
speed which are singular in the small wavenumber limit. We assume, in the
spirit of mode-coupling methods, $B(k)=Bk^{-s}$ (we assume an isotropic scale
dependence for $B(k)$ for simplicity which suffices our purpose of finding the
scale dependence of the effective Alfv\'en wave speed). Clearly, if $-s+1<z$,
the small wavenumber limit of the problem is dominated by underdamped waves, if
$-s+1>z$ the Alfv\'en waves are damped out in the small wavenumber limit. In
contrast if $-s+1=z$ then in the small wavenumber limit both the propagating
and the dissipative modes are present. Our analyses in Sec.\ref{symm} clearly
suggest that if there are fluctuation corrections to $B_o$ then the {\em
effective} scale dependent Alfv\'en wave speed $B(l)\sim k^{-\chi}$ yielding
$s=\chi$. Furthermore, since $\chi+z=1$ we have $-s+1=z$ leading to the
co-existence of the underdamped Alfv\'en waves and the dissipative modes in the
large-scale, long-time limit. Note that this situation allows us to define a
dimensionless parameter $\beta\equiv B/\eta_+$ where $B$ and $\eta_+$ are the
renormalised Alfv\'en speed and the viscosity respectively.

In the SCMC approach vertex corrections are neglected. Lack of vertex
renormalisations in the zero wavevector limit in $3d$MHD  allows SCMC to yield
{\em exact relations} between the scaling exponents $\chi$ and $z$ as in the
noisy Burgers/Kardar-Parisi-Zhang equation \cite{freykpz}. In this model the
nonlinearities and the noise variances do not renormalise, thus
 leading to $z=2/3,\, \chi=1/3$, satisfying the exponent identity $\chi+z=1$.
In the present case, due to the singular nature (in the small wavevector limit)
of the bare noise correlations (\ref{noisecor}), they do not pick up any
further singular corrections to their scaling; however the amplitudes get
modified (this is similar to the results in Ref.\cite{abepl}). 
We denote the renormalized amplitudes by $D$ and $\tilde D$,
respectively. Furthermore, for the $3d$MHD Eqs. (\ref{mhdfullv}) and
(\ref{mhdfullb}) there are infra-red singular fluctuation corrections to $B_o$
[see the one-loop diagrams in Fig. (\ref{diag1})] leading to $B_o(l)\sim
l^{\chi}$ consistent with our arguments above. The SCMC approach involves
consistency in the scaling and the amplitudes of the mode coupling equations.
It should be noted that the exponent values $z=2/3,\,\chi=1/3$ satisfy the mode
coupling integral equations regardless of the strength of the mean magnetic
field. For amplitude consistency the one-loop integrals are required to be
evaluated. Due their complicated structure, we evaluated them by assuming that
the strength of the mean magnetic field is small. Therefore, only our amplitude
relations and not the scaling exponents are affected by the approximation of
small mean magnetic fields.

Demanding amplitude consistency in the mode coupling equations we obtain,
\begin{eqnarray}
\eta_+&=&
\frac{D}{\eta_+^2}\left[1-\frac{2}{d}+\frac{1}{d(d+2)}\right]+\frac{\tilde
D}{d(d+2)\eta_+^2}-\frac{\beta^2 D}{\eta_+^2d}\left[1-\frac{2}{d}+
\frac{1}{d(d+2)}
-\frac{\beta^2 \tilde D}{\eta_+^2d(d+2)(d+4)}\right],\nonumber \\
\beta&\equiv& \frac{B}{\eta^+},\;B=B_o\frac{D}{d(d+2)\eta_+^3},
\nonumber \\
\frac{1-\Gamma}{1+\Gamma}&=&\frac{(1+\Gamma^2) (1-\frac{2}{d}+\frac{2}{d(d+2)})
-0.5\Gamma (1-\frac{2}{d})+F_1(\beta)}{(1+\Gamma^2)
(1-\frac{2}{d}+\frac{2}{d(d+2)})+0.5\Gamma (1-\frac{2}{d})+F_2(\beta)},
\label{ampli1}
\end{eqnarray}
where $F_1(\beta)\equiv 2\beta^2(1+3\Gamma^2)(-1/d+\frac{1}{d(d+2)}- \frac{1}
{d(d+2)(d+4)}),\;F_2(\beta)\equiv 2\beta^2(3+\Gamma^2)
(-1/d+\frac{1}{d(d+2)}- \frac{1} {d(d+2)(d+4)})$ are the $\beta$-dependent
parts of the amplitude-ratio.
While calculating the amplitude consistency relations we worked in the limit of
small $\beta\equiv B/\eta_+$ and the renormalised magnetic Prandtl number
($\frac{\eta}{\nu}=\frac{\eta_+-\eta_-}{\eta_++\eta_-} $ ) close
to unity. Away from these limits the amplitudes of the
underlying one-loop integrals become much
more complicated functions of $\beta$ and also of the renormalised magnetic
Prandtl number, but the qualitative picture remains unchanged.
The main physical
picture that emerges from the expressions (\ref{ampli1}) is that in the
absence of
a bare mean magnetic field $B_o$ the effective Alfv\'en wave speed is also zero
which is a restatement of the fact that if the original theory is isotropic, so
will be the renormalised theory. Moreover the renormalised Alfv\'en speed
 increases with increasing
 $B_o$. With these results we are now in a position to calculate the
energy spectra in the inertial range. We use the form of the effective
(scale dependent) viscosity and the Alfv\'en wave speed to
obtain the equal time correlation function of $z^{\pm}_i$ in the
long wavelength limit. We define (in $3d$)
\begin{eqnarray}
C^{++}_{ij}(k,\omega)&=&\langle
z_i^+(k,\omega)z_j^+(-k,-\omega)\rangle= \frac{2Dk^{-3}P_{ij}({\bf
k})}{(\omega -Bk^{-1/3}k_z)^2+\eta_+^2k^{4/3}},\nonumber
\\
C^{--}_{ij}(k,\omega)&=&\langle
z_i^-(k,\omega)z_j^-(-k,-\omega)\rangle=\frac{2Dk^{-3}P_{ij}({\bf
k})}{(\omega+ Bk^{-1/3}k_z)^2+\eta_+^2k^{4/3}},\nonumber
\\
C^{+-}_{ij}(k,\omega)&=&\langle
z_i^+(k,\omega)z_j^-(-k,-\omega)\rangle=\frac{2\tilde{D}k^{-4/3}P_{ij}({\bf
k})}{
[-i(\omega-Bk^{-1/3}k_z)+\eta_+k^{2/3}][i(\omega+Bk^{-1/3}k_z)+\eta_+k^{2/3}]}.
\label{corrdef}
\end{eqnarray}
As discussed before, we have omitted anisotropic
corrections of $O(B_o)^2$ or of $O(\beta)^2$ to the correlation functions as we
are trying to find out the relevance (in an RG sense) of Alfv\'en waves on the
scaling of the isotropic correlation functions in the absence
of any mean magnetic fields. Therefore, the equal time
correlation functions have the following form (in $d=3$):
\begin{equation}
C^{++}_{ij}(k,t=0)=C^{--}_{ij}(k,t=0)=Dk^{-3-2/3}P_{ij}({\bf k}).
\label{correxp1}
\end{equation}
Therefore, the equal time autocorrelation functions of $\bf
z^{\pm}$ are independent of any mean magnetic field. This holds
true regardless of the scaling of the noise variances.  The equal
time cross correlation function $C^{+-}_{ij}(k,t=0)$ requires more
careful considerations. On integrating $C^{+-}_{ij}(k,\omega)$
over all frequency $\omega$ one obtains (in $3d$)
\begin{equation}
C_{ij}^{+-}(k,t=0)=\frac{\tilde{D}k^{-3}P_{ij}}{iBk^{-1/3}k_z+\eta_+k^{2/3}}\,,
\label{correxp2}
\end{equation}
a form which is valid in the inertial range. It is clear from the expression
(\ref{correxp2}) that both the real and the imaginary parts of
$C_{ij}^{+-}(k,t=0)$ scale as $k^{-3-2/3}$ in the inertial range. Thus the
one-dimensional kinetic- and the magnetic-energy spectra (which are simply
related to the correlators defined in Eqs. (\ref{corrdef})) scale as $k^{-5/3}$
in the inertial range. The emerging physical picture is as follows: We find
from the expression for $C^{+-}_{ij}(k,\omega)$ above that this, as a function
of frequency $\omega$, has maxima at $\omega=\pm Bk^{-1/3}k_z$ and the width at
half-maxima $\sim k^{2/3}$. In contrast, the auto correlations of $\bf z^+,z^-$
are maximum at $\omega=0$ and their widths scale as $\sim k^{2/3}$. Thus in the
long wavelength limit the width and the location of the maxima of
$C^{+-}_{ij}(k,\omega)$ scale in the same way leading to the presence of the
underdamped Alfv\'en waves in the hydrodynamic limit. Therefore, it immediately
follows that the kinetic- and the magnetic-energy spectra, being linear
combinations of the correlators discussed above times appropriate phase
factors, scale as $k^{-5/3}$ {\em even} in the presence of a non-zero mean
magnetic field. It should be noted that we have used a small $\beta$
approximation to arrive at our results for the self-consistent amplitude 
relations.  For a finite $\beta$ one would require
to work with a fully anisotropic form of the correlation functions and obtain
self-consistent relations for scaling and anisotropic amplitudes, a task, which
is analytically challenging, remains to be done in the future. However, based
on our calculations above, the exponent identity $\chi+z=1$ and the Ward
identity suggesting that the renormalised Alfv\'en wave speed should scale as
$k^{-1/3}$ with wavevector $k$ being in the inertial range, we argue that {\em
even} for a finite $\beta$, i.e., a finite mean magnetic field the energy
spectra will scale as $k^{-5/3}$ in the inertial range, a result supported by
many observational evidences \cite{obser}. The self-consistent amplitude
equations will then be anisotropic reflecting the presence of underdamped
Alfv\'en waves in the inertial range. The full correlation matrix will be
anisotropic in the hydrodynamic limit; its eigenvalues have different
amplitudes, but all of them scale the same way. Our confidence on our result
that the scaling of the correlation functions along directions parallel and
perpendicular to the direction of the mean magnetic field is same is derived
from the fact that our exponent values $z=2/3,\,\chi=1/3$ satisfy the one-loop
integral equations regardless of the strength of the mean magnetic field and
are consistent with the Ward identity discussed above.

\section{Kolmogorov's constants}
\label{kolcons}
According to the Kolmogorov's hypothesis for fluid turbulence \cite{k41}, in the
inertial range energy spectrum
$E(k)=K_o\epsilon^{2/3}k^{-5/3}$, where $K_o$, a {\em universal}
constant, is the Kolmogorov's constant and $\epsilon$ is the
energy dissipation rate
per unit mass. Various calculations, based on different techniques by different
groups \cite{krai1,leslie,yakhot,jkbamita} show that $K_o\sim 1.5$ in three
dimensions. Having noted that the energy spectra, even in the presence of a
mean magnetic field scale as $k^{-5/3}$ extensions of Kolmogorov's hypothesis
for $3d$MHD allows one to
define Kolmogorov's constants for the Els\"{a}sser fields: $E_{\pm} (k) =
K_o^{\pm} \epsilon_{\pm}^{2/3}k^{-5/3}$.
Since $\bf z^{\pm}=u\pm b$, we have $\langle z_i^+
({\bf k},\omega)z_j^+({\bf -k},-\omega)\rangle= \langle z_i^- ({\bf
k},\omega)z_j^-({\bf -k},-\omega)\rangle$ and $\epsilon_+=\epsilon_-
=\epsilon_{MHD}$ in
absence of any crosscorrelations between the velocity and the magnetic fields,
we have $K_o^+=K_o^-=K_{MHD}$.
The noise strength $D$ and the rate of energy dissipation per unit mass is
connected by the Novikov's theorem \cite{nov,abepl}:$\epsilon=2D\frac{S_3}{(2
\pi)^3}$. Noting that the energy spectra $E_{\pm}(k)$
of $\bf z^{\pm}$ in the inertial range is given by
\begin{equation}
E_{\pm}(k)=Dk^{-3}/\eta_+k^{2/3},
\end{equation}
where $D$ is the {\em effective} or renormalised noise strength,
we identify
\begin{equation}
K_{MHD}=1.6\left[1+0.7\left(3\Gamma^2-6\Gamma-\beta^2
29/105\right)\right]^{2/3}.
\label{kolm}
\end{equation}
The notable feature of the expression (\ref{kolm}) is that the constant
$K^{MHD}$ depends on the dimensionless parameter $\beta$ which we introduced
before. For $\Gamma=0$, i.e., for no magnetic fields, we find $K_{MHD}=K_o=1.6$
for pure fluid turbulence which is well-within the accepted range of values
\cite{leslie}. Before closing this Sec. we would like to point out that the
presence of multiscaling raises questions about Kolmogorov's constant $K_{MHD}$
being universal: A small but finite intermittency correction (i.e.,
multiscaling) over the simple K41 scaling implies the presence of an {\em
arbitrary} scale which may spoil the universality of $K_{MHD}$. We however
refrain ourselves from getting into this question and adopt a point of view
that regardless of whether or not $K_{MHD}$ remains universal due to
multiscaling, the numerical value of this constant is likely to get affected by
the presence of Alfv\'en waves in the system which is reflected by the
expression (\ref{kolm}).

\section{Possibilities of variable multifractality}
\label{inter}
Experiments and numerical sumulations \cite{abprl,cho} find nonlinear
multiscaling
corrections to the K41 prediction of $\zeta_p^a=p/3$ for the structure
functions
in the inertial range. Until the date, no controlled perturbative calculation
for $\zeta_p^a$ is available.
To account for multiscaling in fluid turbulence, however,
Obukhov \cite{obu} and Kolmogorov \cite{kol2} assumed a log-normal distribution
for dissipation $\epsilon$ to arrive at
\begin{equation}
S_p^v(r)=
\langle |\Delta v|^p\rangle=C_p\overline{\epsilon}^{p/3}r^{p/3}\left(\frac{L}{r}
\right)^{(\overline{\delta}/2)p(p-3)},
\label{log}
\end{equation}
where $\overline{\epsilon}$ is the mean value of $\epsilon$ and
(a bound for
$\epsilon$ in fluid turbulence has been discussed in Ref.\cite{doer}):
\begin{equation}
\langle
\epsilon ({\bf x+r})\epsilon ({\bf x})\rangle\propto \langle (\Delta
v)^6/r^2\rangle\sim (L/r)^{9\overline{\delta}}.
\end{equation}
 For small $\overline\delta$,
$\delta\simeq 9\overline{\delta}$. A standard calculation on the randomly
stirred model yields {\em intermittency exponent} $\delta=0.2$ \cite{jkbamita}
where $\delta=9\overline\delta$,
whereas the best possible estimate from experiments is 0.23 \cite{jkbamita}.
This model, despite having well-known limitations and difficulties
\cite{frisch}, serves as a qualitative illustration of multiscaling.
As in fluid turbulence, in MHD the dissipations $\epsilon_{\pm}$
of the Els\"{a}sser variables
$\bf z^{\pm}$ fluctuate in space and time, and as a result one may define
two intermittency exponents $\delta_{\pm}$ for them. In the present problem
$\delta_+=\delta_-=\delta_{MHD}$. Below we calculate the exponent
$\delta_{MHD}$ in a one-loop expression which will give us an estimate of the
Alfv\'en wave speed-dependent deviation of the scaling of the structure
functions from their simple-scaling values as predicted by the K41 theory. We
closely follow Ref.\cite{jkbamita}. We work with the self-consistent forms for the
self-energies and correlation functions given above along with the
consistency relations for the amplitude-ratios $\Gamma {\rm and}\;\; \beta$.
Following Ref.\cite{jkbamita}, we find the dissipation correlation functions in
$3d$ to be
\begin{equation}
\langle\epsilon({\bf x+r})\epsilon({\bf x})\rangle\simeq
12.4\epsilon_{MHD}^{2}\alpha^2 K_{MHD}^2 \ln \frac{L}{r},
\label{diss}
\end{equation}
with $\alpha$ being defined by the relation $\nu_+=\alpha\epsilon_{MHD}^{1/3}$.
From the self-consistent amplitude-relations (\ref{ampli1}) we find
\begin{equation}
\alpha=0.4\left[1+0.7\left(3\Gamma^2-6\Gamma-\beta^2
29/105\right)\right]^{1/3}\left[1-0.5\Gamma
-4\beta^2(0.7-0.03\Gamma)\right]^{1/3}.
\label{alpha}
\end{equation}
Thus, $\alpha$ which is a universal coefficient in ordinary fluid turbulence,
varies with the parameter $\beta$, or with the Alfv\'en wave speed in MHD.
Substituting the values of $K_o$ and $\alpha$ we find
\begin{equation}
\delta_+=\delta_-=\delta_{MHD}=0.2\left[1+0.7\left(3\Gamma^2-6\Gamma-\beta^2
29/105\right)\right]^{4/3}.
\label{delta}
\end{equation}
Thus we find that a decrease in the value of $\delta$ with an increase in
$\beta$, i.e., with increasing mean magnetic field. Despite the limited
applicabilities of log-normal models in characterising multiscaling, we can
conclude, from our expression (\ref{delta}) for the intermittency exponent in
MHD in the presence of a mean magnetic field, that the intermittency
corrections to the simple K41 scaling is likely to get affected (reduced in our
calculations) in the presence of Alfv\'en waves. Further calculations and/or
numerical simulations are needed to find the exact extent of the dependence of
multiscaling on the mean magnetic field and the possibilities of anisotropic
multiscaling for the structure functions parallel and perpendicular to the mean
magnetic field as demonstrated recently in Ref.\cite{biskampnew}. It should be
noted that our conclusions on the multiscaling properties of the structure
functions depend on a log-normal model for $3d$MHD. Such a description,
unfortunately, is unable to distinguish between the structure functions
parallel and perpendicular to the direction of the mean magnetic field in the
system. Moreover, the intermittency exponents above [expressions (\ref{delta})]
are evaluated in the lowest order in mean magnetic field. Therefore, even
though from our results we are not able to make firm comments on the
possibility of parallel and perpendicular structure functions exhibiting
different multiscaling, the real importance of our results lie in their
elucidation of the multiscaling properties depending on the mean magnetic
field.

\section{Conclusion}
\label{conclu}

In this paper we have considered the effects of the Alfv\'en waves on the
statistical properties of the correlation functions of the velocity and the
magnetic fields or the Els\"{a}sser fields. We considered the case when the
velocity fields are incompressible. In a one-loop approximation we find that
the {\em effective} or the renormalised Alfv\'en wave speed scales as
$k^{-1/3}$ where the wavevector $k$ is in the inertial range. This immediately
yields that the energy spectra, even in the presence of a mean magnetic field,
scale as $k^{-5/3}$ in the inertial range. We identify a dimensionless
parameter $\beta$ which is the ratio of the effective Alfv\'en wave speed and
the renormalised viscosity. We obtain self-consistent relations between the
amplitude ratio $\Gamma$ of the correlation functions and $\beta$. These
relations allow us to calculate the dimensionless Kolmogorov's constant and we
show that it depends explicitly on $\beta$ or on the mean magnetic field.
Finally, we calculate the intermittency exponent which in a log-normal model
gives a qualitative account of the multiscaling in terms of the deviation from
the K41 scaling for the structure functions. We would like to emphasize that
although the one-dimensional Burgers-like model of MHD of Ref.\cite{abjkb} and
the $3d$MHD Eqs. yield energy spectra independent of the Alfv\'en waves, the
long wavelength physical pictures are different. In the former case, due to the
nonrenormalization of the bare Alfv\'en wave speed, Alfv\'en waves are
overdamped in the hydrodynamic limit; the dominant process in that limit is
viscous dissipation. In contrast, in $3d$MHD, the Alfv\'en wave speed picks up
a singular correction in the hydrodynamic limit leading to the K41 scaling of
the energy spectra and the presence of underdamped Alfv\'en waves in the
hydrodynamic limit. Despite similar mathematical structures these crucial
differences arise principally because the one-dimensional model decouples
completely when written in terms of the Els\"{a}sser variables, allowing to
comove with the waves of each of them separately. As a result correlators of
each of them are independent of the Alfv\'en waves and the bare Alfv\'en wave
speed remains unrenormalised leading to overdamped Alfv\'en waves in the
hydrodynamic limit. However, the $3d$MHD equations do not decouple when written
in terms of the Els\"{a}sser variables and hence oppositely propagating waves
cannot be made to vanish by comoving. Despite the limitations of the one-loop
methods \cite{yakhot} and the small $\beta$ approximation to facilitate easier
analytical manipulations for the amplitude relations, we obtain results which
are significantly new and open the intriguing possibilities of MHD multiscaling
universality classes being parametrised by the mean magnetic field. Some of the
quantitative details will change if one retains terms which are higher order in
$\beta$; however, we believe that the qualitative picture will essentially
remain the same. As MHD turbulence forms a natural example of a driven
nonequilibrium system with Alfv\'en waves, our results are the first of its
kind for a natural system. In our log-normal model approach, we did not
distinguish between the longitudinal and the transverse structure functions our
results cannot be compared directly with those of Ref.\cite{biskampnew}, where
the longitudinal and the transverse structure functions are found to scale
differently with multiscaling exponents which depend on the magnitude of the
mean magnetic field in the presence of a finite magnetic helicity. In contrast
our results apply to the multiscaling of the usual structure functions in the
absence of any magnetic helicity which are combinations of the transverse and
the longitudinal structure functions as considered in Ref.\cite{biskampnew}
which would also then exhibit a mean magnetic field dependent scaling. In
particular we find that the usual structure functions multiscale less in
presence of an increasingly strong mean magnetic field which is in qualitative
agreement with those of Ref.\cite{biskampnew}. Although here we have restricted
ourselves to the study of Alfv\'en waves as relevant perturbations on the
amplitude and the scaling of the isotropic correlation functions in the limit
of small $\beta$, our results indicate the possibility of the multiscaling
exponents varying with the amplitude of the mean magnetic field. Our analyses,
when extended for finite values of $\beta$ and for full anisotropic structure
of the correlation functions, are likely to provide further understanding of
and resolve some of the discrepancies between the various phenomenological
scenarios available for situations with large mean magnetic fields
\cite{galtier,ng,montgomery}. We leave this task for the future.

From a broader point of view our results demonstrate the critical role of
wave-like excitations in determining the statistical properties of $3d$MHD. The
presence of propagating waves is not confined to MHD only; they are a generic
feature in many other naturally occurring soft-matter systems where such waves
can be present in a viscous environment, e.g., active polar gels in
cytoskeletal dynamics \cite{frank} and in the dynamics of self-propelled
particles \cite{sriram} etc. We believe our results will lead to similar
theoretical and experimental studies in relevant nonequilibrium soft-matter
systems.

\section{Acknowledgement}
This work was started when one of the authors (AB) was an Alexander von
Humboldt Fellow at the Hahn Meitner Institut, Berlin.
AB wishes to thank the AvH Stiftung, Germany for partial financial support.

\end{document}